\documentclass{appolb}
\usepackage{graphicx}

\begin{document}
 \eqsec  
\title{Pulse Shape Discrimination in Liquid Argon and its Implications for Dark Matter Searches Using Depleted Argon%
\thanks{Presented at the Epiphany 2012 Conference}%
}
\author{Pawel Kryczynski on behalf of the WArP R\&D Group
\address{The Henryk Niewodniczanski Institute of Nuclear Physics Polish Academy of Sciences, Krakow}
\\
\address{pawel.kryczynski@ifj.edu.pl}
}
\maketitle
\begin{abstract}
A brief outline of Dark Matter detection experiments using liquid Argon technology is presented. The Pulse Shape background discrimination method (PSD) is described and the example of its use in 2.3 l R\& D detector is given. Methods of calculating sensitivity of a Dark Matter detector are discussed and used to estimate the possible improvement of sensitivity after introduction of isotopically depleted liquid Argon. 
\end{abstract}
\section{Dark Matter Detection}
\subsection{Experimental evidence and candidates}
Since the beginning of the 20th century the number of experimental facts suggesting the insufficiency of using solely luminous matter in the description of the Universe has been growing. Several measurements have shown that the baryonic matter provides only the 4.5\%  of the mass - energy of the Universe. Of the remainder: about 73\% is associated with Dark Energy (Cosmological Constant) and over 22\% with Dark Matter \cite{pdg}. 

One of the prominent observations suggesting the existence of DM was the measurement of the shape of the rotation curves of galaxies.
\begin{figure}[htb]
\centering
\includegraphics[width=0.48\textwidth]{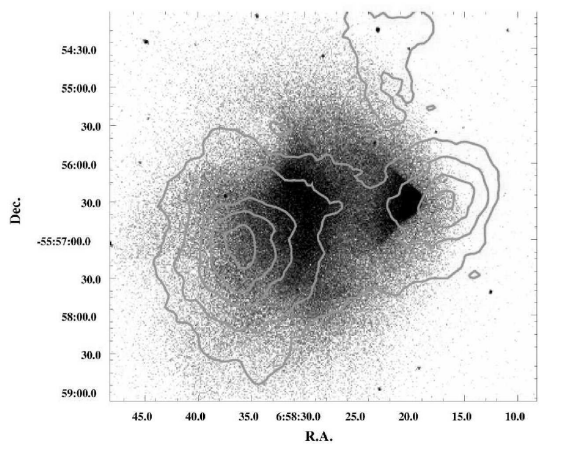} \\
\caption{One of the evidences for the Dark Matter existence: the Bullet Cluster \cite{bullet_cluster}.}
\label{Fig:F2H}
\end{figure}
\begin{figure}[htb]
\centering
 \includegraphics[width=0.6\textwidth]{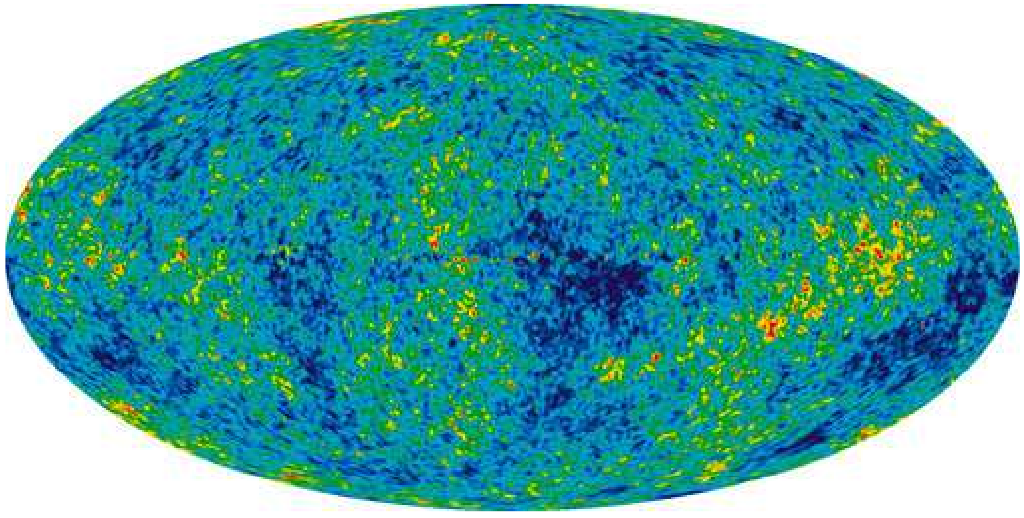} \\
\caption{Another important evidence for the Dark Matter existence: the map of the CMB distribution from the WMAP experiment \cite{cmb}.}
\label{Fig:F2Hm}
\end{figure}
The stars rotate around the center of the galaxy with such velocities as if the mass was distributed in an uniform way, and not in the central bulge and the surrounding disk. The observation of so called "polar ring galaxies" \cite{polar_ring} (with additional stars rotating perpendicularly to the main disk) leads to a similar conclusion.

Similar conclusions can be drawn from the observations of objects such as the Bullet Cluster (Figure ~\ref{Fig:F2H} right) where the movement of the luminous parts does not trace that of the main mass component of the colliding clusters as measured by weak lensing. The measurements of Cosmic Microwave Background Radiation (CMB), Figure ~\ref{Fig:F2Hm}, also support the hypothesis of Dark Matter. The anisotropies in the CMB can help calculate of the percentage of the baryonic matter in the Universe and the share of the Dark Matter in its total mass.

Today, most astrophysicists agree that Dark Matter exists - it is its nature that is the question. The number of candidate particles is great. These can be divided into warm DM (the relativistic particles which freezed out at the early stages of the Universe formation), which will not be discussed here, and the cold DM which includes particles with non - relativistic velocities and yet not known form. The last is included in the preferred cosmological model - $\Lambda$CDM. The most popular and promising DM candidate is a WIMP (a Weakly Interacting Massive Particle). Its allowed mass range extends from tens of GeV to above 1 TeV \cite{wimp}. The excluded cross sections are of over $10^{-44}$ cm$^2$ for spin independent interactions and over $10^{-39}$ cm$^2$ for spin - dependent interactions \cite{Fox:2012ee}. As the velocities are non relativistic, the expected energies of recoiling nuclei in a detector are of the order of tens of keV  \cite{sl}. Candidates for WIMPs may be supersymmetric particles (e.g neutralino). The expected event rate is very low - 0.02 events/day/kg of detector mass. The issue of Dark Matter detection has to be dealt with by employing efficient detecting techniques and and reducing the background to unprecedented levels.

\subsection{Overview of current detection techniques}
\begin{figure}[htb]
\centerline{%
  \includegraphics[width=0.55\textwidth]{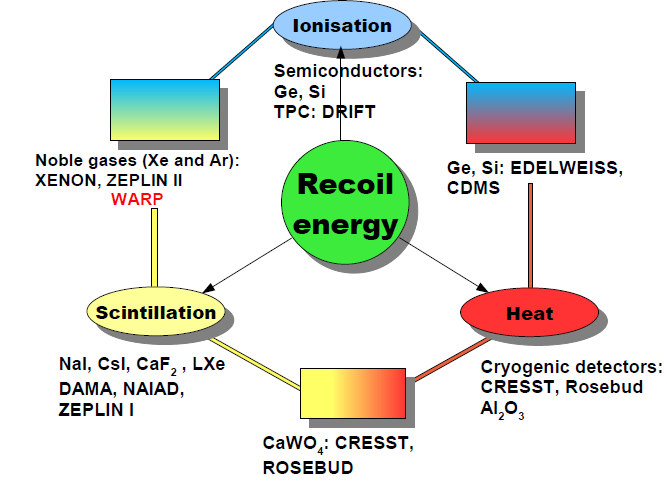}}
\caption{Classification of experiments according to their detection techniques \cite{ep2006}}.
\label{Fig:F3H}
\end{figure}

The methods of Dark Matter detection can be divided into indirect and direct. The former uses products of annihilation or decay of the DM particles and is not discussed here, whereas the latter is connected with the observation of DM interactions in a detector. The transfered energy can be detected as ionization of the detector medium, heat in the crystal lattice of the detector or scintillation (examples of experiments are given in Figure ~\ref{Fig:F3H}). Typically, a combination of two of the mentioned signals is used to obtain a better background discrimination. Cryogenic detectors using Ge or Si crystals (CDMS) detect ionization and heat, detectors based on CaWO$_{4}$ use the scintillation signal and heat and liquid noble gas detectors - ionization (when working in double phase mode) and scintillation.

\section{Liquid Argon as a Dark Matter detector}
The analysis presented in this work uses some of the methods developed in \cite{canci} and expands on the results presented in \cite{robert} and as such, will focus on liquid Argon detectors. A typical double-phase Argon DM detector consists of a dewar filled with liquid argon with photomultipliers observing a fiducial volume. Over the liquid, there is a layer of gaseous Ar.

\begin{figure}[htb]
\centerline{%
  \includegraphics[width=0.24\textwidth]{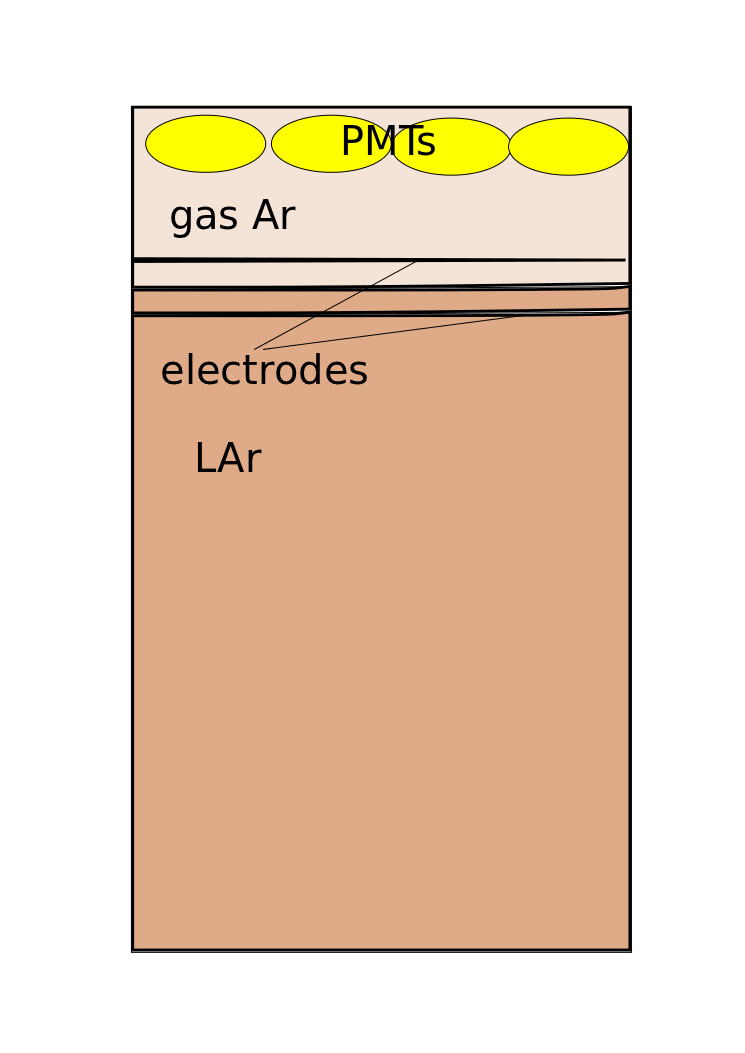}}
\caption{A scheme of a LAr DM detector (based on \cite{scheme}).}
\label{Fig:F41H}
\end{figure}

The detector can work in single phase mode, registering only the light signal produced in the liquid (S1). In this detection mode the only signal which is measured is the scintillation. When a DM particle hits Argon nucleus, it may become excited or ionized and form a dimer Ar$_{2}$. The relaxation of the excited dimer may be of two forms - singlet (fast signal) with decay time about 7 ns and triplet (slow signal) with decay time 1500 ns. Such a significant difference (absent in Xenon, also used in DM detection) offers a background discrimination possibility because electrons produce slow and fast components in a different proportion than neutrons (and WIMPs). This results in different signal shapes. When the double phase mode is used, the ionization caused by the recoiling nuclei is extracted to the gas layer above the liquid by the electrodes situated below and over the surface to cause the secondary signal (S2) proportional to the ionization (Figure ~\ref{Fig:F41H} ).

\begin{figure}[htb]
\begin{tabular}{ c c }
  \includegraphics[width=0.45\textwidth]{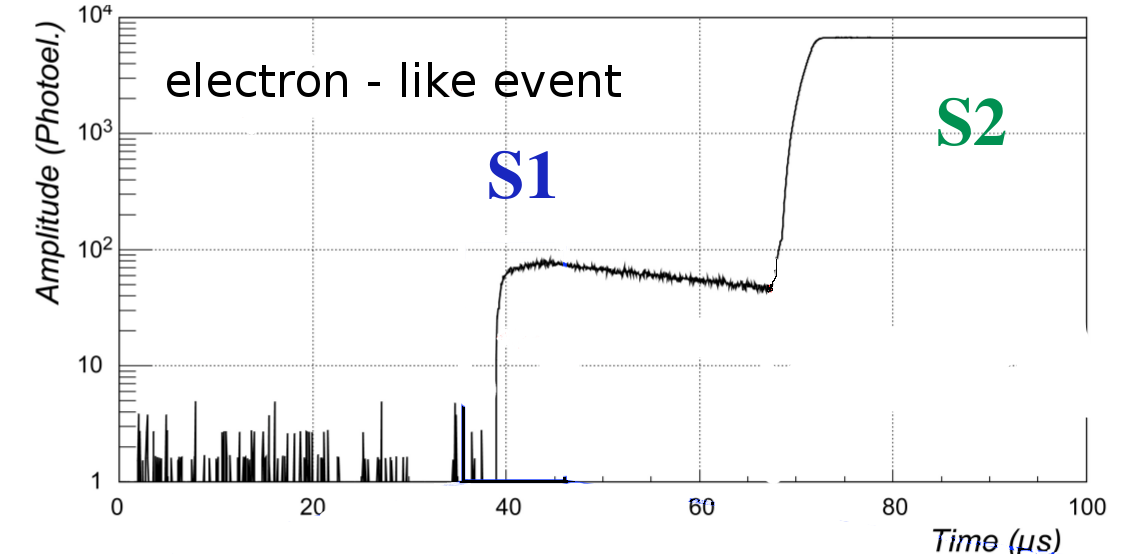} & \includegraphics[width=0.45\textwidth]{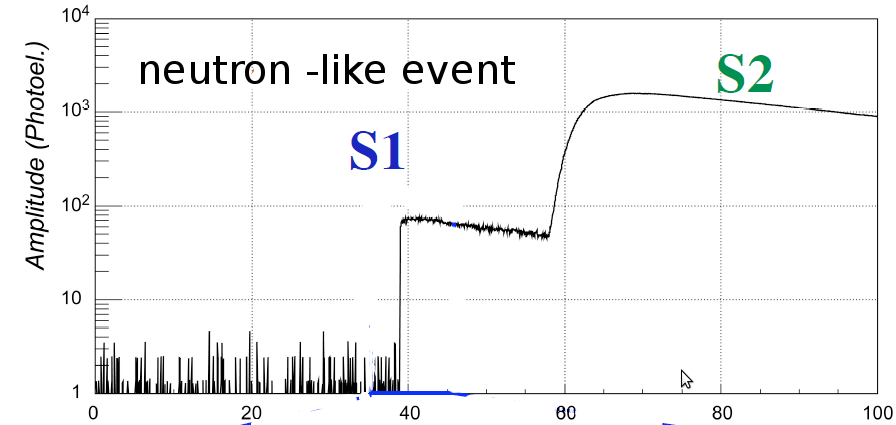}  \\
\end{tabular}

\caption{A difference of the S1 and S2 signals for an electron - like event (left) and the neutron - like event (right) \cite{canci}.}
\label{Fig:F71H}
\end{figure}

Using S1 and S2, the difference in the ratio of relative amplitudes of signals resulting from electron-like events (main background) and neutron (or WIMP)-like events is an additional discrimination method, resulting in a better rejection of background events (Figure ~\ref{Fig:F71H}).
\subsection{Background suppression}
The background in DM detectors can be divided into two subclasses: external and intrinsic. Both are caused mainly by $\gamma$ rays and neutrons. The effects of the background resulting from the cosmic rays (dominating during tests performed on the earth surface) and the natural radioactivity of the rocks (in underground laboratories), can be reduced by precise measurements and simulations leading to determination of their signature as well as by using shielding. 

The intrinsic background connected with the construction of the detector can be diminished by using radiopurified materials, but the most important part of it is related to the radioactive isotopes $^{39}$Ar and $^{42}$Ar present in atmospheric Argon along with non radioactive $^{40}$Ar. Produced in interaction of cosmic rays or due to the neutron capture, these isotopes contribute to the background due to $\beta^-$ decays \cite{argon39}. The activity of $^{39}$Ar dominates and was measured to be 1.01 $\pm$ 0.02(stat) $\pm$ 0.08(syst) Bq for 1 kg of liquid Argon \cite{ar39}.

To reduce background effects, in double phase detection mode the S1/S2 discrimination method can be applied. In single phase measurements, like presented in this work, the Pulse Shape Discrimination has to be applied. We have used the FPrompt method described below, as it offers good discrimination power and its implementation is not difficult. The studies which led to the choice of FPrompt are described in \cite{robert}. The parameter used to separate background and signal is calculated according to the formula:
\begin{equation}
F_{P}=\frac{S_{F_{P}}}{S_{1}}=\frac{\int^{T_{F_{P}}}_{T_{i}}V(t)}{\int^{T_{F}}_{T_{i}}V(t)}
\end{equation}
where $T_i$ and $T_F$ are the limits of the time window for signal $V(t)$ and $T_{F_P}$ is the time which ensures the best separation of two populations - 100 ns for analysis of 2011 data and 120 ns for the data from 2009 \cite{robert}.

\subsection{Detector upgrade performed in 2010/2011}
\begin{figure}[htb]
\centering
\begin{tabular}{ c c c }
  \includegraphics[width=0.35\textwidth]{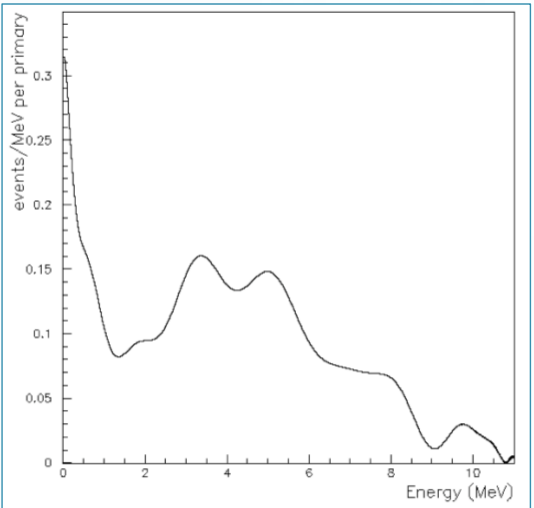} & \includegraphics[width=0.38\textwidth]{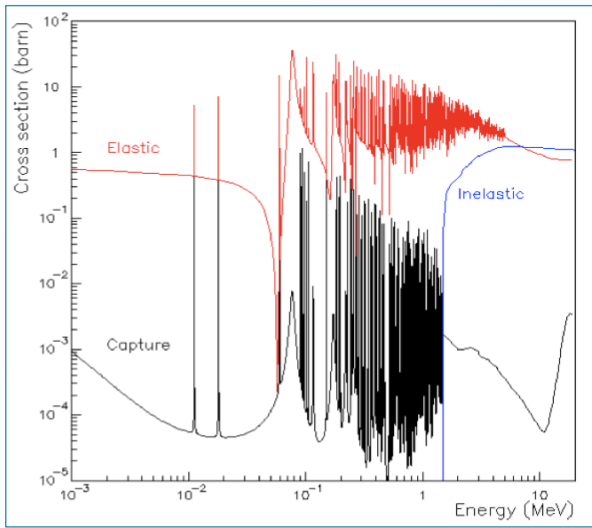} \\
\end{tabular}

\caption{The spectrum of the Am/Be source (left) \cite{ambe} and references therein and the cross section for neutron interaction on argon nuclei (right) \cite{canci} and references therein.}
\label{Fig:F10H}
\end{figure}

\begin{figure}[htb]
\begin{tabular}{ c c }
  \includegraphics[width=0.17\textwidth]{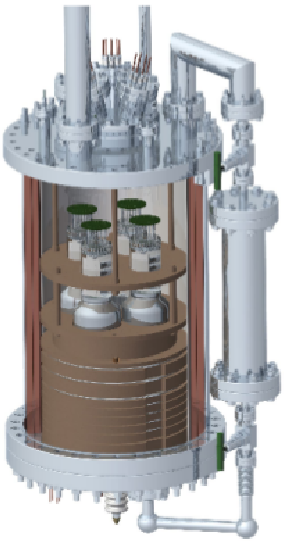} &  \includegraphics[width=0.6\textwidth]{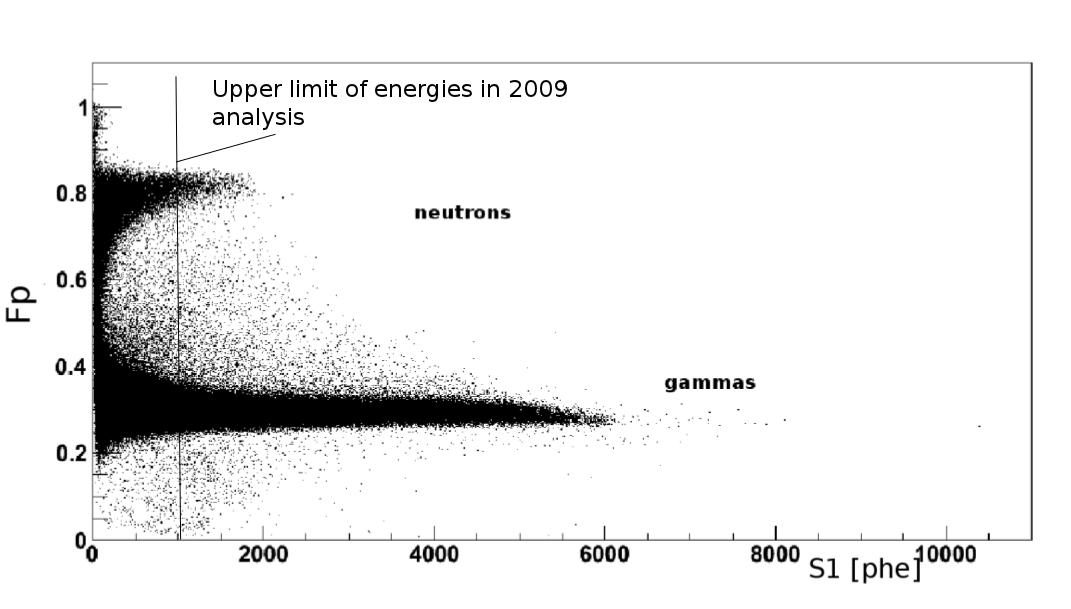} \\
\end{tabular}

\caption{The the detector scheme (left) \cite{andrzej} and a difference of the energy ranges in 2009 and 2011 data (right).}
\label{Fig:F72H}
\end{figure}

The WArP 2.3 liter prototype detector (Figure ~\ref{Fig:F72H} left) was the first to report DM search results in liquid Argon \cite{benetti} and has been since used for R~\&D purposes. In 2009 the WArP collaboration measured the $n$-like ev./$e^{-}$-like ev. separation by irradiating a chamber with an Am/Be source. In the FPrompt distributions calculated using this data, an intermediate population was found. It was connected with the inelastic scattering of neutrons and included in the model which was then fitted to the spectrum.  Indeed, such scattering dominates for energies of the particles emitted by the Am/Be source (Figure ~\ref{Fig:F10H}). The data suffered from a low light yield and due to some DAQ problems it was not possible to subtract the background. After a detector upgrade, we decided to repeat this measurement profiting from a four times higher LY. This upgrade, performed in 2010, is described in detail in \cite{upgrade1} and \cite{upgrade2}. New photomultipliers with higher quantum efficiency were installed, which led to a higher light yield (6.1 phel/keV compared to 1.52 phel/keV in 2009 runs). The consequence of this was an insight into a region of lower recoil energies. The new DAQ electronics with a broader dynamic range, causing less saturated events, made the exploration of higher energies possible (Figure ~\ref{Fig:F72H} right).
\subsection{Analysis of 2011 data and reanalysis of 2009 data}
\begin{figure}[htb]
\centering
  \includegraphics[width=0.45\textwidth]{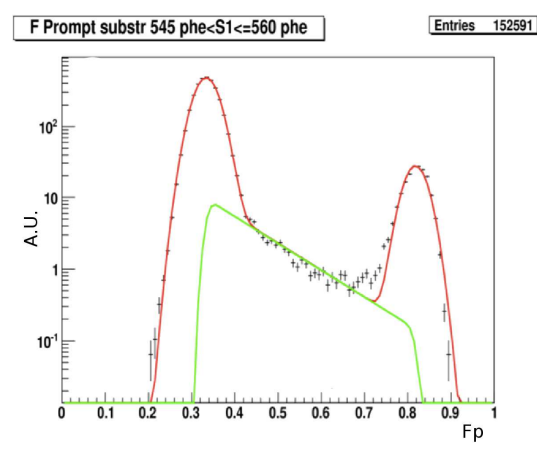}
\caption{The FPrompt spectrum with an intermediate population visible ( this analysis results).}
\label{Fig:F10Hm}
\end{figure}

In 2011, runs were performed to test the new detector setup. The Am/Be neutron source was used to imitate the WIMP interactions and the $^{241}$Am to perform detector calibration and to estimate the light yield. The results from 2009 were used as a reference. The FPrompt spectrum was fitted with a sum of two Gaussian functions (for gamma and neutron population) and a convolution of an exponential and Gaussian function (for the intermediate - inelastic neutron interaction population):
\begin{equation}
G_{n}\bigoplus (G_{i}\bigotimes E_{i}) \bigoplus G_{\gamma}
\end{equation}
where the $G_{n}$ and $G_{\gamma}$ are the Gaussian distributions for a neutron and gamma population respectively and $G_{i}\bigotimes E_{i}$ is the convolution.
 One of the purposes of this analysis was to test whether the intermediate population would be observed (one of the energy bins is presented in Figure ~\ref{Fig:F10Hm}).
 
\begin{figure}[htb]
\centering
  \includegraphics[width=0.6\textwidth]{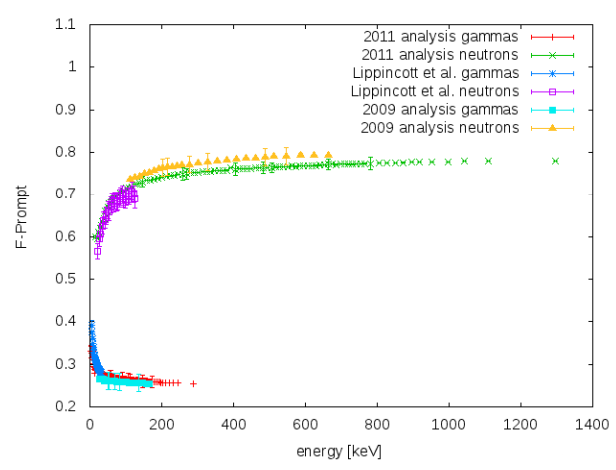} 
\caption{Comparison of the FPrompt spectrum for the 2011 and 2009 WArP data \cite{canci} and the results published by Lippincott et. al \cite{lippincott}.}
\label{Fig:F11H}
\end{figure}

The positions of the gamma and neutron populations resulting from these fits are compared with the previous measurements and published results from other groups (\cite{robert}, \cite{lippincott}) in Figure ~\ref{Fig:F11H}.

\begin{figure}[htb]
\centering
  \includegraphics[width=0.5\textwidth]{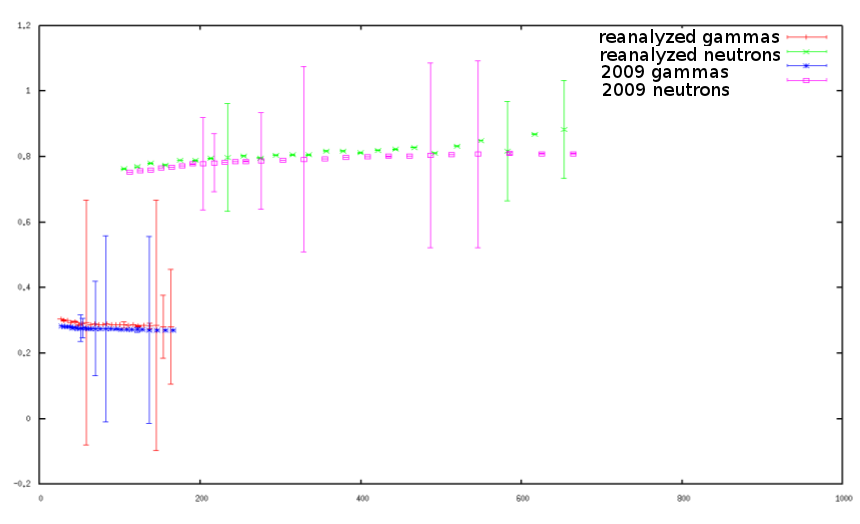} 
\caption{The results of the re-analysis of the 2009 data with the new reconstruction.}
\label{Fig:F12H}
\end{figure}

\begin{figure}[htb]
\centering
  \includegraphics[width=0.6\textwidth]{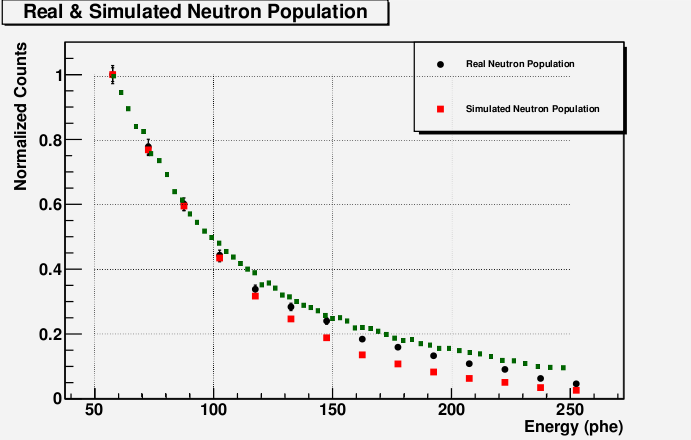} 
\caption{Comparison of the neutron population share in the fitted FPrompt spectrum. The black dots is the 2009 data, the red - the MC simulation (figure from \cite{robert}) and the green - the 2011 data without background subtraction applied added to the plot.}
\label{Fig:F13H}
\end{figure}

In order to understand the small discrepancy between the new data and the 2009 results, a re-analysis of the 2009 data with the use of new scripts and reconstruction procedures was performed. The preliminary results are presented in Figures ~\ref{Fig:F12H} and ~\ref{Fig:F13H}. Figure ~\ref{Fig:F13H} presents  a comparison of the number of elastic neutron interactions per unit of energy to the number expected from  Monte Carlo simulations.

\section{Predicted Sensitivity for a Depleted Argon Detector}
\subsection{Calculating the Sensitivity of a Dark Matter Detector}
Even if a DM detector sees no signal, it is still possible to set the exclusion limits on the allowed $\sigma$ vs WIMP mass parameter space. Typically, the model accounting for the Earth galaxy motion, the velocity distribution of WIMPs in the galactic halo, seasonal DM flux change and detector efficiency is applied, as described e.g in \cite{sl}. The resulting function predicting the number of observed events as a function of recoil energy is integrated taking the detector threshold into account. Repeating this procedure for all WIMP masses and taking into account the detector exposure leads to the sensitivity plot. The tools using this approach were prepared and successfully tested on published data. Another way to calculate detector sensitivity was proposed by S. Yellin \cite{yellin} and is based on the distribution of events as a function of their energy. This was also tested but not used in presented analysis.

\subsection{Depleted Argon}
One of the crucial parameters in calculating the sensitivity of a DM detector is its exposure. Although the sensitivity should increase with the exposure, it is limited due to the presence of radioactive $^{39}$Ar isotope. Its abundance causes the pile - up of the signal and leakage of the gamma population (background) into the signal region. The isotopically depleted Argon seems to be a promising solution. The abundance of the $^{39}$Ar can be reduced by depletion in centrifuges but costs of such production method are high. The other approach is to search for underground Argon, which may be found e.g in reservoirs of Helium and then distilled. Using this method, a  depletion factor of 25 has been confirmed \cite{depleted}.
\subsection{Predicted sensitivity for a Depleted Argon detector}
\begin{figure}[htb]
\begin{tabular}{ c c }
  \includegraphics[width=0.4\textwidth]{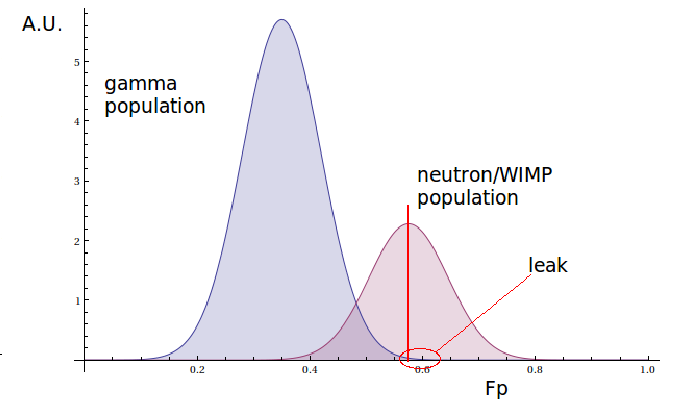} & \includegraphics[width=0.45\textwidth]{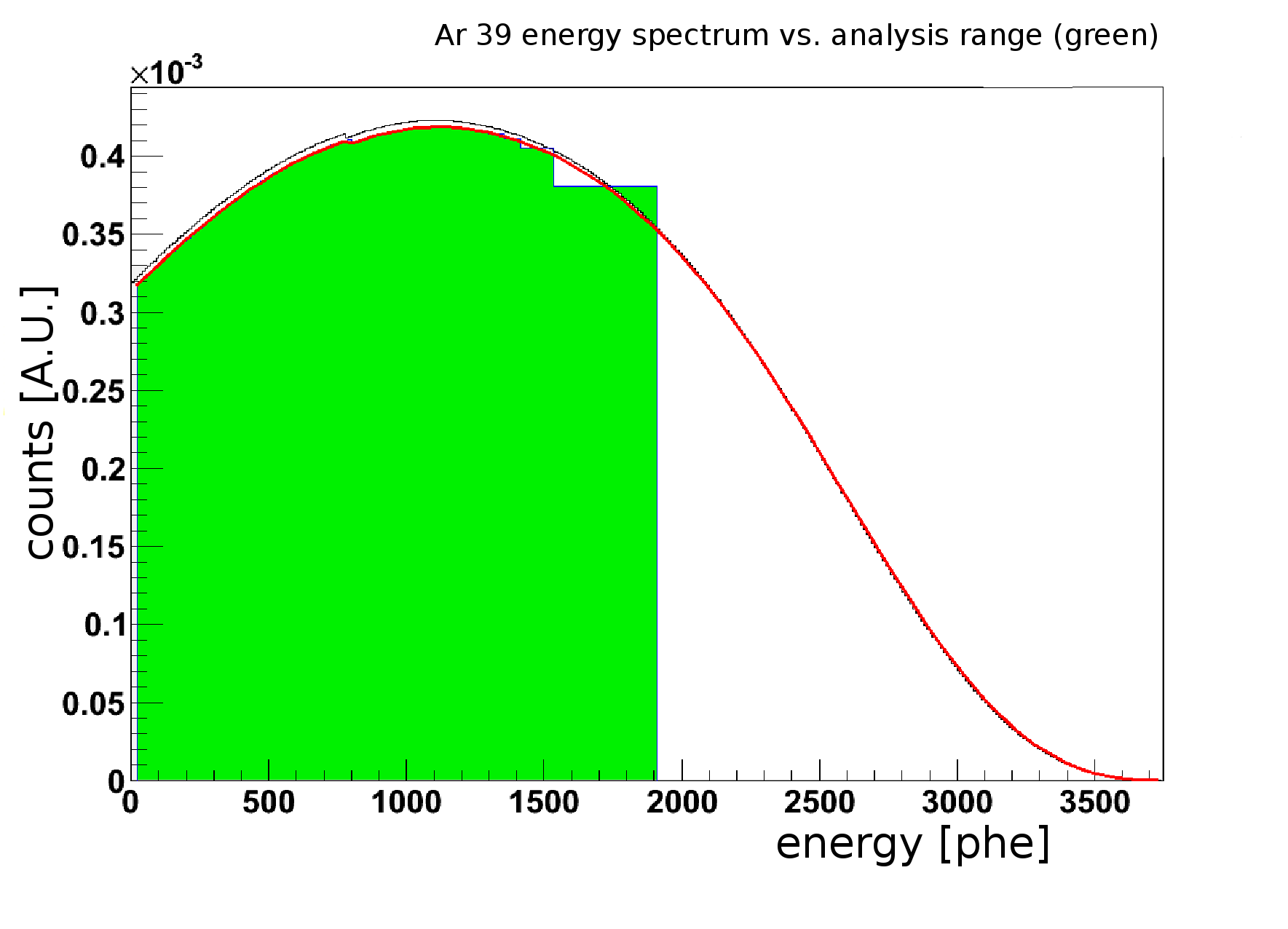}  \\
\end{tabular}

\caption{The method used to estimate exposure (left) and the $^{39}$Ar spectrum with the analysis range marked in green (right).}
\label{Fig:F14H}
\end{figure}

The method used in this work is based on the known $^{39}$Ar energy spectrum and radioactivity. The maximum exposure before the background events start to "leak" into the conservatively chosen signal region is calculated. To do this, the events are randomly chosen from the analysis energy range and according to the known parameters of the Gaussian function in the corresponding bin the number of entries in the signal window is estimated. This, thanks to the known radioactivity of $^{39}$Ar, helps to estimate the possible exposure (Figure ~\ref{Fig:F14H}).

\begin{figure}[htb]
\centerline{%
\includegraphics[width=0.5\textwidth]{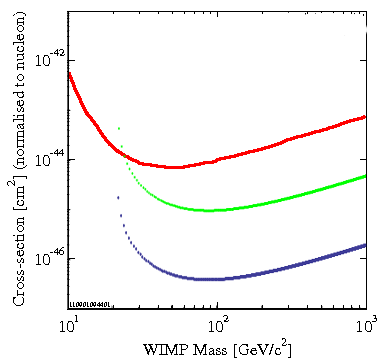}}
\caption{The comparison of sensitivity obtainable in a liquid argon detector with (blue) and without (green) the depleted Argon. Results from XENON 100(2011) \cite{xenon2011} were used as the reference (red curve) and plotted using tool \cite{dmplotter}. Other curves are obtained using tools prepared for this analysis.}
\label{Fig:F15H}
\end{figure}

The calculations of the predicted exposure have been performed assuming $^{39}$Ar to be the only background. The results from XENON (2011) were used as the reference and are presented in Figure ~\ref{Fig:F15H} (\cite{dmplotter}, \cite{xenon2011}). The case with the S2/S1 discrimination applied (with conservative assumption of the background rejection power of $10^2$) is displayed with a green curve and the case with a depleted Argon (by factor 25) - with a blue curve.

\section{Conclusions}
The preliminary results for the analysis of the data obtained with the upgraded detector setup were presented. The insight into a wider energy range has been reported. The comparison with the data from 2009 has been performed. The influence of the possible use of the isotopically depleted Argon has been demonstrated and the significant possible improvement of sensitivity has been seen.
\section*{Acknowledgements}
I would like to thank the WArP R \&D group for providing me with the data used in this work and for the help with my analysis. In particular, I would like to thank A. Szelc both for the support during the data analysis and for correcting this article. I want also to thank A. Zalewska for reading and correcting this work.
\section*{•}

\end{document}